\documentclass{article}
\usepackage{xcolor}
\usepackage{bm,ulem}
\usepackage{amsmath}
\usepackage{graphicx} 
\usepackage{authblk}
\usepackage[colorlinks=true, linkcolor=blue, urlcolor=blue,
  citecolor=black]{hyperref}

\usepackage{amssymb}
\usepackage{lineno}
\usepackage{epstopdf, epsfig}
\usepackage{appendix}

\newcommand{\beq}{\begin{equation}}
\newcommand{\eeq}{  \end{equation}}
\newcommand{\beqa}{\begin{eqnarray}}
\newcommand{\eeqa}{  \end{eqnarray}}
\usepackage{amsmath}

\usepackage{authblk}
\graphicspath{{../}}

\title{The dominant role of jetting in micron- and sub-micron sea spray produced by bubble bursting: a revised model and comparison with measurements.}

\author[1,2,*]{Alfonso M. Ga{\~n\'a}n-Calvo}
\author[1]{Miguel A. Herrada}
\author[1]{José M. López-Herrera}

\affil[1]{Dept. Ing. Aerospacial y Mec{\'a}nica de Fluidos,Universidad de Sevilla.\\
Camino de los Descubrimientos s/n, 41092 Sevilla, Spain.}
\affil[2]{ENGREEN, Laboratory of Engineering for Energy and Environmental Sustainability, Universidad de Sevilla, 41092 Sevilla, Spain.}

\begin{document}
\maketitle
*email: amgc@us.es

\begin{abstract}
The primary physical mechanism governing the production of sub-micron sea spray aerosols (SSA) --specifically the competition between film and jet droplets from bubble bursting-- has remained a subject of intense debate. This work presents a revised, first-principles model that establishes the overwhelming dominance of jetting for producing micron and sub-micron aerosols. Our approach first rules out the film droplet mechanism as a primary contributor for this size range by demonstrating through physical scaling and numerical simulations that the final average ejected volume of sub-micron droplets is much smaller than that from jetting. A comprehensive global probability distribution function (PDF) for SSA is constructed by rigorously modeling its fundamental components in sequence: (i) refining the sub-surface bubble size distribution with a simpler and better experimentally supported exponential law, and (ii) deriving consistent number and size distribution models of droplets per bursting event from an ample set of high-resolution numerical simulations. The droplet size PDF from a single bubble follows a highly-skewed distribution --optimally modeled by a Generalized Inverse Gaussian distribution-- revealing the production of nanometric droplets previously unaccounted for by simpler models. When integrated, these components yield a final predictive model for the global SSA size distribution, with parameters derived directly from physical principles and simulations rather than empirical fitting. The model demonstrates extraordinary predictive accuracy, aligning almost perfectly with experimental data from both laboratory and oceanic measurements, in particular across the critical 25 nm to 2.5 $\mu$m range. This research significantly enhances the fundamental understanding of marine aerosol generation and provides a more accurate foundation for climate and atmospheric chemistry models.
\end{abstract}

\section{Introduction}

Marine aerosols are critically influential in the Earth's climate system and the global water cycle serving as cloud condensation nuclei (CCN) and ice nucleation particles (INP). These aerosol particles significantly influence cloud microphysical processes, affect atmospheric chemistry, modulate Earth's radiative balance, and consequently play a fundamental role in climate regulation \cite{Andreae2008,Leeuw2011}. Despite their recognized importance, the exact origins and precise mechanisms responsible for generating ultra-fine marine aerosols have remained topics of intense scientific debate and research efforts\cite{Wang2017,Mayer2020,JRVW22,VWD22,G23,LG24}. These aerosols are classified as primary sea spray aerosols (SSA) and secondary marine aerosols (SMA). In particular, Mayer et al. \cite{Mayer2020} have established a largely accepted and reinforced cornerstone concept in marine atmospheric chemistry by convincingly showing the dominance of the secondary pathway under the conditions of their simulated phytoplankton bloom. Current literature is refining this conclusion to state that the balance between the primary organic-rich SSA and secondary SMA (from VOCs) pathways is a key variable that likely changes with location, season, and the specific biology of the ecosystem. In contrast, salt-rich SSA remain a stable but less dominant overall concentration level. Thus, the debate has shifted to understanding the relative balance between the primary and secondary pathways, which varies by region and ecosystem. However, the Mayer et al. study and subsequent work have made understanding SSA production even more crucial since SSA provides the seeds for SMA: both nucleation and condensation mechanisms necessitate preexisting particles whose composition change along the process.

In marine atmosphere, there are almost always pre-existing SSA particles from bubble bursting. These particles provide the pristine surface for the newly formed secondary material (sulfate, organics) to stick to. Therefore, the number, size, and surface area of the SSA particles produced by bubble bursting directly control (i) whether new particles will form from available gas molecules, (ii) how the secondary mass will be distributed across the aerosol population, and (iii) the final size and chemical composition of the CCN. In addition, while SMA may be the key driver of changes in the number of cloud-forming particles, larger salt-rich SSA particles completely dominate the total mass and surface area of marine aerosol. This is important for other climate processes like the direct radiative effect: These larger particles are very effective at scattering sunlight back to space, which has a direct cooling effect on the planet, independent of clouds. Moreover, the vast surface area of sea salt particles provides a reactive medium for chemical reactions in the atmosphere (e.g., reactions involving nitrogen and halogen compounds) that can alter atmospheric composition\cite{Patterson2016}. Consequently, the precise description of aerosol production from bubble bursting and the quantification of the resulting aerosols is a hotter area of research than ever. Therefore, quantifying the SSA spectra is not just about understanding one source; it is a prerequisite for correctly predicting the evolution and impact of the entire marine aerosol system. In particular, precisely determining the origin of the sub-micron aerosol population is indeed crucial, as these particles represent the primary vector of CCN.

A substantial controversy persists within the literature regarding whether the primary source of sub-micron aerosols arises from film droplets produced by the rupture of bubble films at the ocean surface \cite{Lewis2004,LV12}, or from jet droplets generated via the dynamic instability of collapsing bubble cavities \cite{Spiel1995,Veron2015}. Classical studies have described both mechanisms, yet achieving definitive conclusions has been impeded by the inherent complexity in accurately characterizing the statistical behavior of bubble populations and their interaction with turbulent ocean surfaces \cite{BW80,Thorpe1992,Deane2002,Deike2016,Mostert2021,Deike2022}.

Recent attempts to model ocean ultrafine spray size distributions resulting from bubble bursting were presented by \cite{Berny2021,G23}. In the case of \cite{G23}, the model demonstrated a critical shortcoming, as it required a probably unrealistically large number of droplets per bubble event to achieve acceptable agreement with experimental observations. Without this inflated droplet count, the predictions deviated significantly from measured distributions. The fundamental cause of this discrepancy stemmed from a simplified assumption, linking the average droplet diameter exclusively to the size of the initially ejected droplet. Despite this limitation, \cite{G23} challenged the widely accepted idea that bubble jetting produces droplets larger than 1 µm and that it contributes negligibly to the generation of CCN. In the opposite direction, the authors of \cite{JRVW22,VWD22} entertained the idea that film flapping could resolve the paradox — or ``anomaly'', as stated by Blanchard and Syzdek in \cite{BS88} — by extending the possibilities that film breakup, incorporating a flapping mechanism stemming from the finite density of air, could provide to our understanding of ultrafine (submicron) aerosol and CCN generation from bubble bursting.

However, some considerations readily exclude film breakup as significant contributors to the droplet size spectrum of interest:

1- The simple equilibrium shape of bubbles of unperturbed radius $R_o$ at the seawater surface is determined by the Bond number ${\rm Bo}=\frac{\rho g R_o^2}{\sigma}$, where $\rho$ and $\sigma$ are the liquid density and surface tension of the liquid, and $g$ is the acceleration normal to the free seawater surface. For small values of Bo, the volume of the liquid film contained by van der Waals forces is several orders of magnitude smaller than that of the liquid ejected by bubble jetting for bubbles smaller than 1 mm of radius at a free air-water surface.

2- The Taylor-Culick velocity argument by \cite{JRVW22,VWD22}, uses a residence time of the rim on the cap that greatly overestimates the actual size of the cap for these small bubbles, which is a small fraction of the bubble radius, $R_o$. For this reason, together with the above, the over-predicted number of film droplets by bubbles with radii below 0.4 mm, as described by \cite{VWD22} in their Figure 4 -- was already questioned in \cite{G22PNAS}. Once the negligible volume of the film cap compared to the jetting volume and the significant number of sub-micron droplets produced by jetting compared to film breakup for $R_o$ below $\sim\, 400 \mu$m is demonstrated in the next sections, film droplets will be finally discounted in this bubble size range.

3- The viscous-capillary length scale, $l_\mu=\mu^2/(\rho \sigma)$, where $\mu$, $\rho$ and $\sigma$ are the liquid viscosity, density and surface tension, respectively, yields $l_\mu\simeq 19.2$ nm in seawater at an average temperature ocean temperature of 15.5$^o$C. The scales of small bubble films that could flutter at the order or smaller than $l_\mu$ are about one order of magnitude smaller than the molecular mean free path in air. It it unlikely that the continuous mechanism necessary to develop the film "flapping" observed at larger scales can be achieved at those small scales.

4- The three-dimensional, high-resolution simulations of small bubbles approaching a free surface and bursting presented in this work indicate that the presumed ``flapping'' mechanism is absent. Additionally, the possibility of the accidental generation of nanodroplets at the instant of film puncturing is reduced to just one or two isolated droplets, whose relative velocity to the free surface is orders of magnitude smaller than that of jetting nanodrops.

Consequently, the proposal by \cite{JRVW22,VWD22} not only do not resolve Blanchard \& Syzdeck's ``anomaly'', but also leave just one potential contributor to the missing nanodroplet count, which also provides its necessary energetic shooting: the bubble jetting. This possibility was recently substantiated by high-resolution simulations in \cite{LG24}, which proved the productivity of elusive droplets as small as 0.1 times $l_\mu$ by jetting, which aligns well with established principles of liquid fragmentation \cite{EV08}. Despite their brief existence due to the axial coalescence in axisymmetric simulations, a number of these droplets are shown to survive at the end of the transient ejection.

Now, to address the limitation in \cite{G23} around the number of droplets and refine the predictive capabilities of aerosol generation models, our study introduces a more nuanced characterization of droplet size distributions generated by bubble jetting. In this work, we reuse the data from \cite{LG24} and perform a statistical analysis of the ejected droplet size distribution. As a result, we obtain a probability density function (PDF) that is highly left-skewed with a mean compatible with the established classical value of $0.1 R_o$. However, bubbles with radii below 0.2 mm (and above 10 µm) eventually produce significantly more nanometer-size droplets than those with sizes proportional to 0.1 $R_o$. Specifically, these smaller droplets approach dimensions consistent with the capillary-viscous length scale of seawater --approximately 19.2 nm at the typical ocean temperature of 15.5 $^o$C. Thus, this new PDF offers a new perspective on the problem without the need to resort to a hypothetical production a large number of droplets proportional to the scale of the first ejected droplet \cite{G23}. In this regard, the new PDF is not a product of fitting to the data, but a result obtained from numerical simulations.

Incorporating this refined distribution has led to a drastically improved prediction that align closely with the experimental observations documented in the existing literature cited in \cite{G23}, whose data are used again here. In summary, our analysis explicitly demonstrates that film breakup droplets contribute insignificantly to aerosol populations within the micron and sub-micron size ranges critical for cloud formation (CCN and INP) and the global hydrological cycle. These film droplets, characterized by volumes several orders of magnitude smaller than those generated through jetting, thus have negligible influence on atmospheric aerosol populations essential for cloud microphysics. Our research has yielded advanced insights that definitively resolve longstanding uncertainties and clearly emphasize the dominance of jet droplets as a primary source of sub-micron marine aerosols (SSA). Given the critical importance of aerosols at the scale of CCN and INP, this clarification enhances fundamental scientific understanding and significantly benefits climate modeling accuracy, potential resuspension mechanisms of pathogens and micro- and nanoplastics, and the formulation of environmental policy and mitigation strategies.

\section{The building blocks of a global spray size probability distribution function (PDF)}

The marginal PDF $P(r_d)$ of the spray size produced by bubble bursting, where $r_d$ is the droplet radius, can be expressed in terms of the combined PDFs of the bubble size distribution $q(R_o/\langle R_o\rangle)$ bursting on the ocean's surface and the droplet size distribution $f(r_d/\langle r_d\rangle,R_o)$ produced by the bursting of bubbles of size $R_o$. It can be expressed as \cite{LV12,Berny2021,G23}:
\begin{equation}
P(r_d)= \int_{0}^{\infty} q(R_o/\langle R_o\rangle) f(r_d/\langle r_d\rangle,R_o) N_d(R_o) {\rm d}R_o.
\label{marginalPDF}
\end{equation}
where $N_d(R_o)$ is the average number of droplets produced by each bursting event. In this expression, it should be noted that both $q$ and $f$ are PDFs normalized with their respective first moments (in contrast to \cite{LV12}). This PDF can be normalized by $l_\mu$ introducing the normalized variables $\chi_d=r_d/l_\mu$ and ${\rm La}=R_o/l_\mu$, which yield
\begin{equation}
P(\chi_d)=\int_{0}^{\infty} q(\text{La}/ \langle \text{La} \rangle) f\left(\chi_d/\langle \chi_d\rangle,{\rm La} \right) N_d({\rm La})\,  d \text{La},
\label{P}
\end{equation}
Thus, three items are needed to complete the sought-for PDF, $P(\chi_d)$: $q$, $f$ and $N_d$.

\subsection{Sub-surface and bursting bubble size distribution}

\begin{figure*}
\centerline{\includegraphics[width=0.80\textwidth]{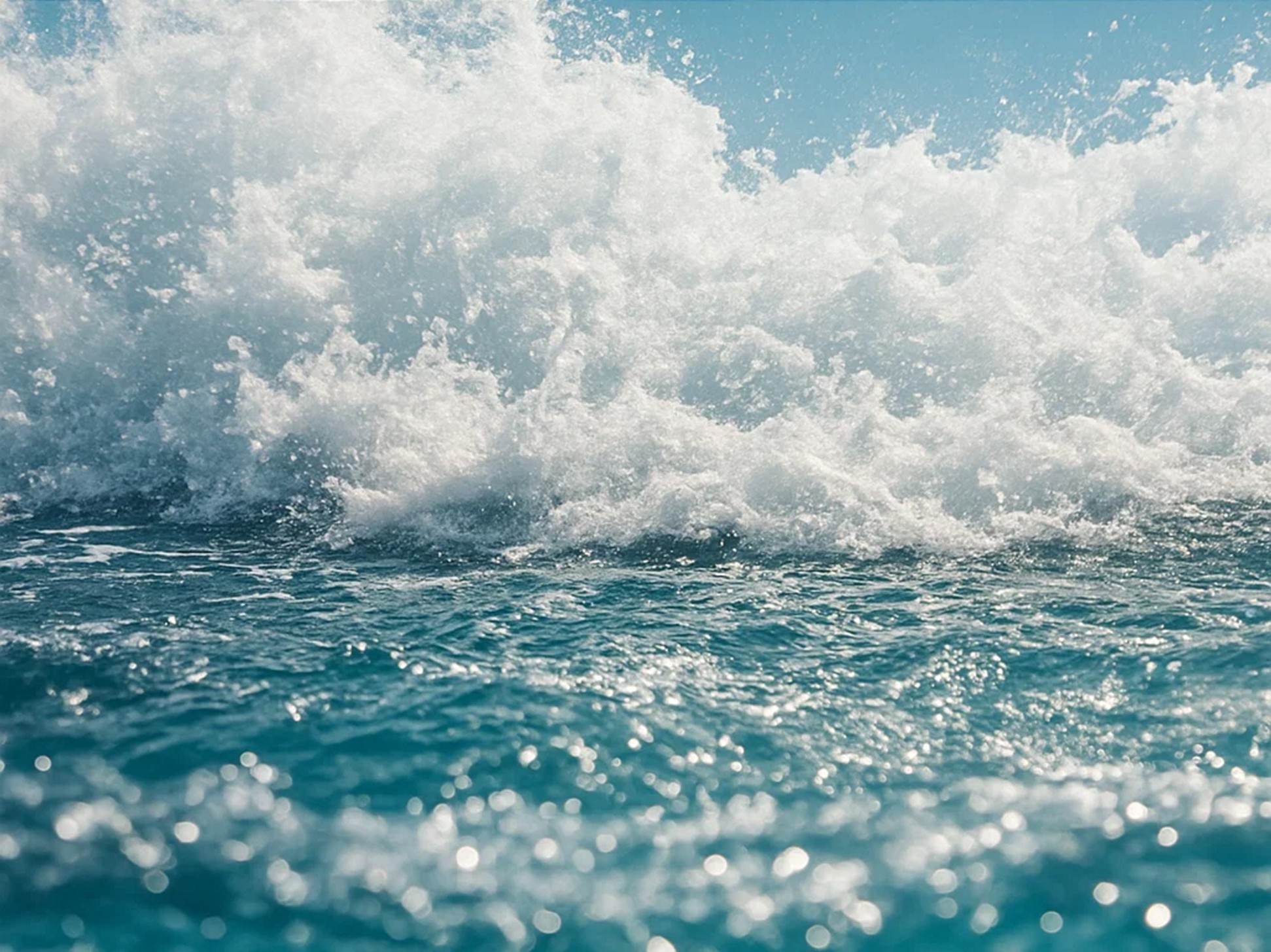}}
\centerline{\includegraphics[width=0.80\textwidth]{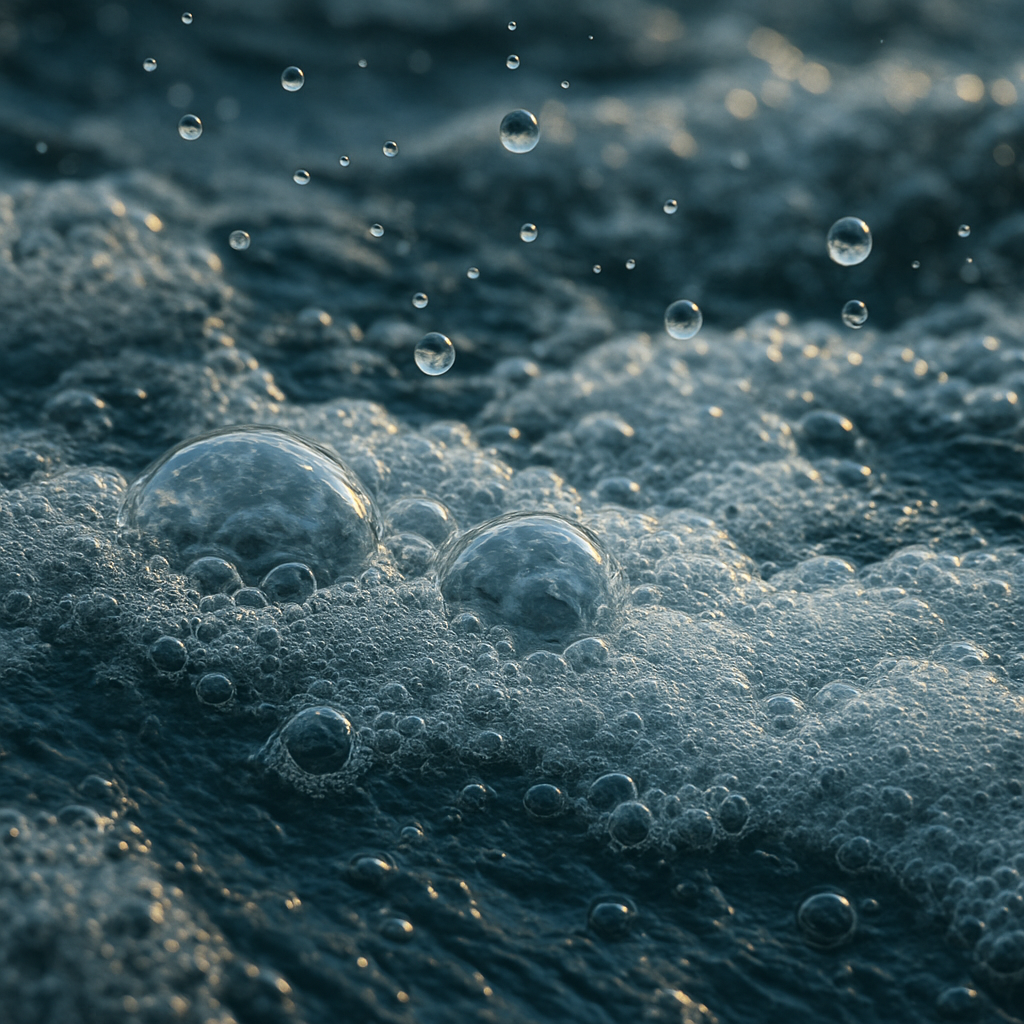}}
  \caption{(Up) Wave breakup snapshot. (Down) An instantaneous snapshot of dynamic ocean spume. Observe the presence of tiny bubbles living at the base rim of the hemispherical caps of large bubbles.}
\label{f1}
\end{figure*}

Breaking waves is the standard mechanism by which subsurface bubbles and spume are produced at the ocean surface. Figure \ref{f1} illustrates the well known wide size spectrum of these bubbles that eventually burst, either isolated or collectively \cite{Neel2021}.

\begin{figure*}
\centerline{\includegraphics[width=0.99\textwidth]{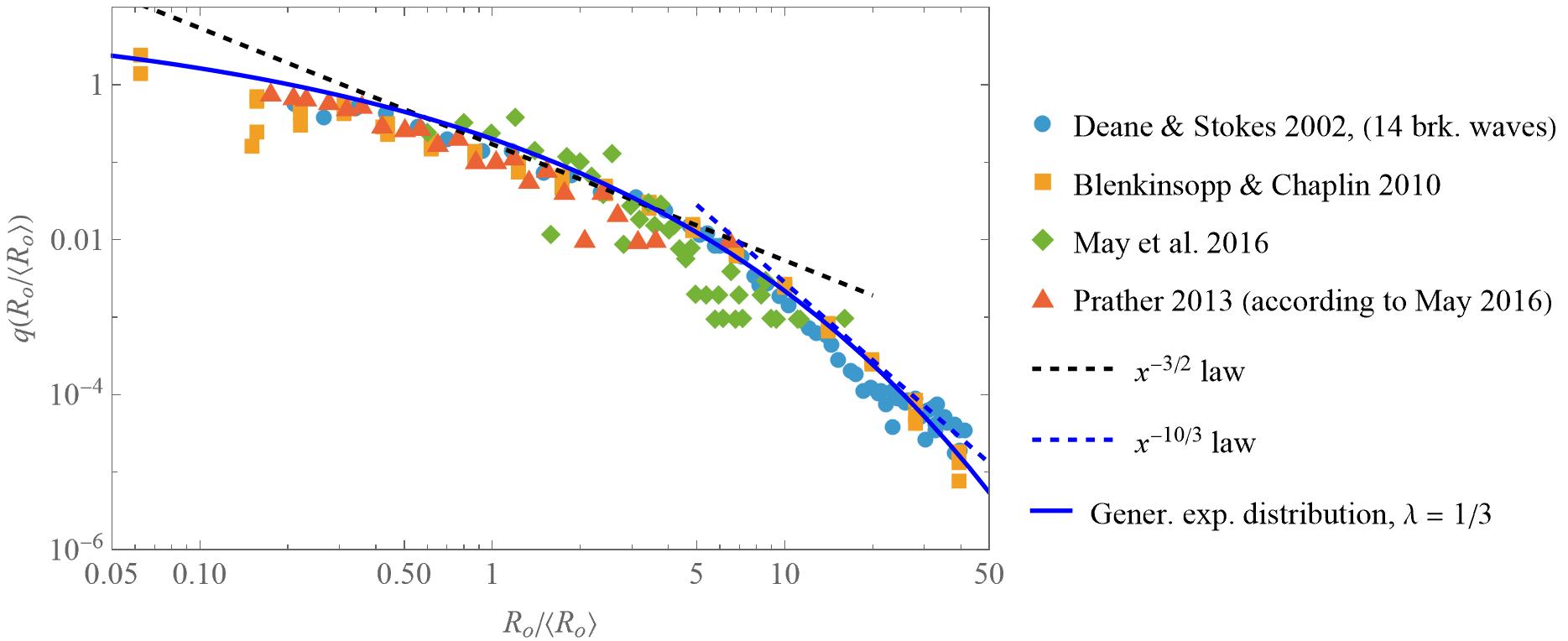}}
  \caption{Bubble size distribution by breaking waves.}
\label{f2}
\end{figure*}

A well established and relatively standard bubble size distribution is that by Deane and Stokes \cite{Deane2002}, which has served as quantitative grounds to further models of SSA size distribution  \cite{Berny2021,G23}. The scaling laws for sub-Hinze and super-Hinze bubble size ranges are $x^{-3/2}$ and $x^{-10/3}$, respectively \cite{Deane2002}, where $x=R_o/\langle R_o \rangle$ and the mean bubble size $\langle R_o \rangle \simeq 0.25$ mm. The power laws are visible in Figure \ref{f2} and agree well with experimental measurements \cite{Deane2002,Blenkinsopp2010,Prather2013}.

However, a simple and even better globally fitting PDF is an exponential distribution with expression
\begin{equation}
q(x)=\frac{\Gamma \left(\frac{2}{\lambda }\right) e^{-\left(\frac{x \, \Gamma \left(\frac{2}{\lambda }\right)}{\lambda  \Gamma \left(1+\frac{1}{\lambda }\right)}\right)^{\lambda }}}{\lambda  \Gamma \left(1+\frac{1}{\lambda }\right)^2}
\label{exp}
\end{equation}
where the best fitting is obtained with $\lambda\simeq 0.33$, as one can observe in Figure \ref{f2} (continuous blue line). This PDF reflects the lack of memory in a Markovian process in which the variable $x^\lambda$ is generated. Interestingly, in this case, $\lambda$ is not 3, as one might naïvely expect from the bubble volume, but rather about 1/3. One possible reason is that the value of $\lambda$ that makes the exponential PDF follow the two limiting power laws proposed by Deane (2002) more closely is precisely 1/3. However, this is not the case for $x < 0.5$. Here, the exponential distribution fits the experimental measurements much better than the $x^{-3/2}$ power law. This could be expected for a memoryless process with small values of $x$: In this range, the probability is nearly constant, independently of $\lambda$. A deeper insight into this issue is beyond the scope of this work, but this PDF will be the one used for subsequent analysis.

\subsubsection{Bursting bubble PDF: Where do sub-millimeter and micrometric bubbles arrive, and what accelerations do they undergo?}

There is still a lack of consensus in whether the measured subsurface size spectrum agrees with that actually bursting per unit time at the surface \cite{Yu2023}. However, a number of evidences support that, indeed, that spectrum does not deviate from the one measured and represented by (\ref{exp}):

1- Recent work by Czerski et al. \cite{Czerski20221,Czerski20222} indicates that bubble size distributions in the sub-Hinze range do not significantly vary with distance to the surface under a given wind speed condition within the first two meters of the sub-surface. This suggests a collective rise mechanism, whereby smaller bubbles are drawn into the wakes of larger bubbles.

\begin{figure*}
\centerline{\includegraphics[width=1.0\textwidth]{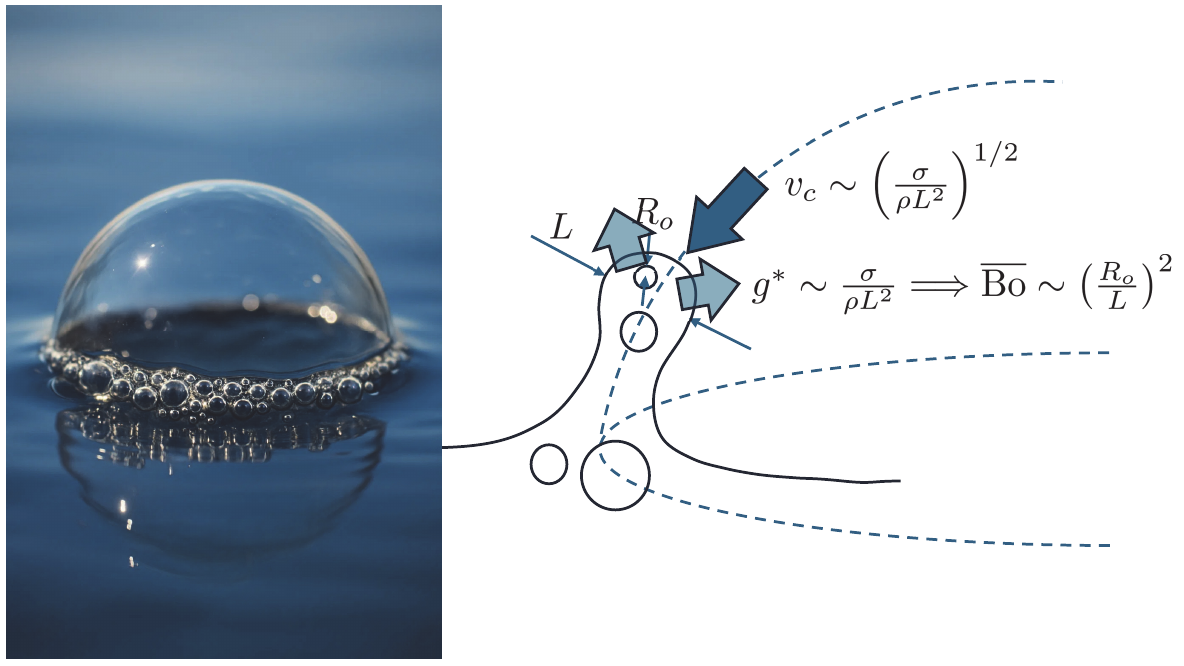}}
  \caption{A large bubble holding a number of much smaller droplets in the rim at its base --a so-called {\it hen-and-chick} configuration-- will lead to impulsive local accelerations $g^*$ acting on these small bubbles after the large bubble bursts.}
\label{Hen}
\end{figure*}

2- When the turbulent bubble plume reaches the surface, the collective size spectrum of the spume exposed to air should reflect the same PDF by the above mechanism (see Figure \ref{f1}), but not only this: the spatial distribution of the bubbles promoted by surface tension favors the bursting mechanisms much more effectively that the simple random exposure of the population of bubbles to the water-air interface under gravity alone. For example, observe that the rims at the base of larger bubbles (those who undergo the gravity most by their large Bond number ${\rm Bo}=\frac{\rho g R_o^2}{\sigma}$) and the interstitial volumes between those larger bubbles are filled with bubbles of smaller and tiny sizes\cite{Pugh1996}. This could be termed a {\it hen-and-chick} phenomenon eventually hold by surface tension (see Figure \ref{Hen}). Therefore, when these larger {\it hen} bubbles burst, they do so first due to their greater exposure to perturbations. The characteristic Taylor-Culick speed and acceleration of their film rims are given by
\begin{equation}
v_{\rm TC}=\left(\frac{\sigma}{\rho h}\right)^{1/2},\quad a_{\rm TC}=\frac{\sigma}{\rho h^2}
\end{equation}
where $h$ is the film or interstitial volume thickness, and will directly affect the smaller {\it chick} bubbles that the {\it hen} bubbles hold in their proximal liquid volumes. Here, $h$ can attain a wide range of values depending of the region of the {\it hen} bubble that is affected by film retraction. Thus, one should expect that a small bubble in the proximity to a liquid-air interface in these circumstances would be exposed to sudden peak accelerations
\begin{equation}
g^* \sim  \frac{\sigma}{\rho L^2},
\label{gravity}
\end{equation}
where $L$ would be a characteristic length smaller than the {\it hen} bubble size but larger than its own {\it chick} size. This {\it hen-and-chick} gathering phenomenon is not associated to a predetermined size: on the contrary, it can be observed to hold for a significant size range, as illustrated in Figures \ref{f1} and \ref{Hen}. Consequently, the actual Bond number denoted by $\overline{\rm Bo}$ affecting each bubble size range $R_o$ can be expressed as
\begin{equation}
\overline{\rm Bo}=\frac{\rho g^* R_o^2}{\sigma}=(R_o/L)^2.
\end{equation}

3- The typical local accelerations $g^*$ are sustained during times of the order of $t_L\sim \left(\frac{\rho L^3}{\sigma}\right)^{1/2}$. This latter time is sufficient for the small bubble to approach the surface, since the characteristic time of the latter, $t_o=\left(R_o/g^*\right)^{1/2}$, should be small compared to $t_L$:

\begin{equation}
\frac{t_o}{t_L}=\frac{\left(R_o/g^*\right)^{1/2}}{\left(\frac{\rho L^3}{\sigma}\right)^{1/2}}=\left(\frac{R_o}{L}\right)^{1/2}
\label{timecomparison}
\end{equation}

Obviously, $L$ will always be larger --or at most of the same order-- than $R_o$.

4- It can be hypothesized that $\overline{\rm Bo}$ will typically follow a homogeneous probability distribution across sizes $L$ and $R_o$ with $R_o<L$, dropping for $R_o\sim L$ (e.g. and exponential distribution). This phenomenon is predicted to result in a relatively homogeneous exposure of bubbles across the entire spectrum to air-liquid free surfaces, leading to their bursting. For example, for $L=250\, \mu$m, $g^*$ would reach astonishing values larger than $10^3$ m/s$^2$, resulting in actual Bond numbers smaller than 0.1 for all bubbles below 80 $\mu$m. It is important to note that, despite the expectation that the parameter $\overline{\rm Bo}$ should be small, it must nevertheless be orders of magnitude larger than Bo for sub-millimeter bubbles.

In summary, the bursting rate of the entire population of bubbles is not expected to modify the size distribution of sub-surface bubbles, which will remain consistent with that of bubbles which burst at any given moment. This hypothesis will be validated by the comparison of the marginal PDF obtained with the ocean spray size measurements.

\subsection{Bubble bursting: film versus jetting volumes}

\subsubsection{Cap film volume}
\label{homogeneousbursting}

\begin{figure*}
\centerline{\includegraphics[width=1.0\textwidth]{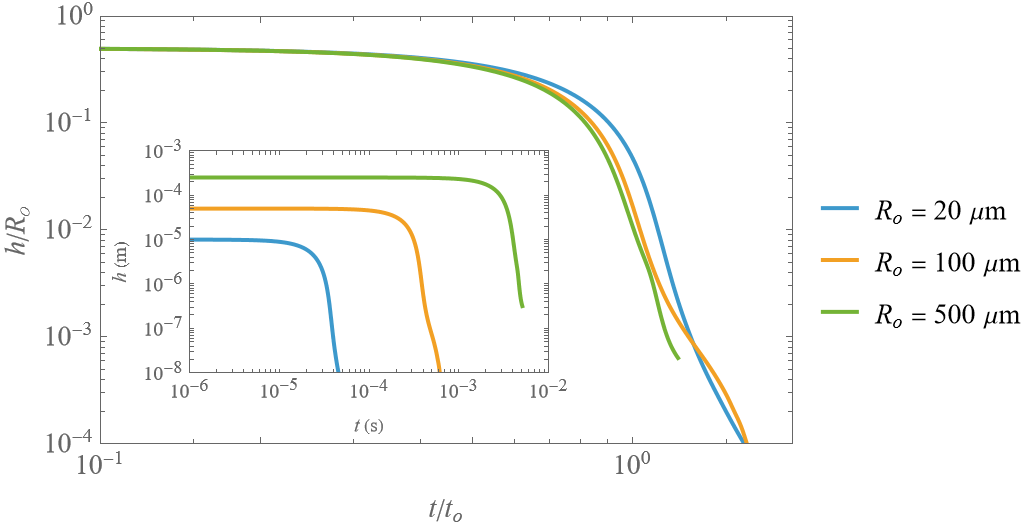}}
  \caption{Draining of bubble caps under the same actual Bond number $\overline{\rm Bo}=0.1$ (sudden acceleration), which verifies the scaling of $t_o$ used.}
\label{f21}
\end{figure*}

When the bubble reaches the water-air interface under a given acceleration $g^*$ given by (\ref{gravity}), the system undergoes a process where the bubble cap exposed to air drains under the action of viscous, inertia and surface tension forces. For small actual $\overline{\rm Bo}=\frac{\rho g^* R_o^2}{\sigma}$ values, and assuming Laplace numbers ${\rm La}=R_o/l_\mu$ larger than ${\rm La}_{cr}\simeq 10^3$ (where ${\rm La}_{cr}$ is the critical Laplace number \cite{Duchemin2002,DGLDZPS18,GL21}), the inviscid draining times of bubbles close to an interface associated to an acceleration proportional to $g^*=\frac{\sigma}{\rho L^2}$ would be comparable to $t_o=\left(R_o/g^*\right)^{1/2} \sim \left(\frac{R_o \rho L^2}{\sigma}\right)^{1/2}\sim \left(\frac{\rho R_o^3}{\sigma \overline{\rm Bo}}\right)^{1/2}$, an augmented  capillary time below milliseconds for $R_o$ below 100 $\mu$m and $\overline{\rm Bo}$ above 0.01. Figure \ref{f21} shows the evolution of the ``cap" size $h(t)$ (distance from the closest point of the bubble to the free surface) of three characteristic bubble sizes ($R_o=20$, 100 and 500 $\mu$m) in seawater, placed at a distance from the free surface $h(0)=0.5 R_o$. The simulations are made with the ultra-precise method {\it JAM} \cite{HM16,JAM} Here, we consider $L=2\times(h(0) + R_o)$. Observe that the proposed scaling correctly collapses the three very different dimensional evolutions due to the impulsive acceleration (inset).

\begin{figure*}
\centerline{\includegraphics[width=0.80\textwidth]{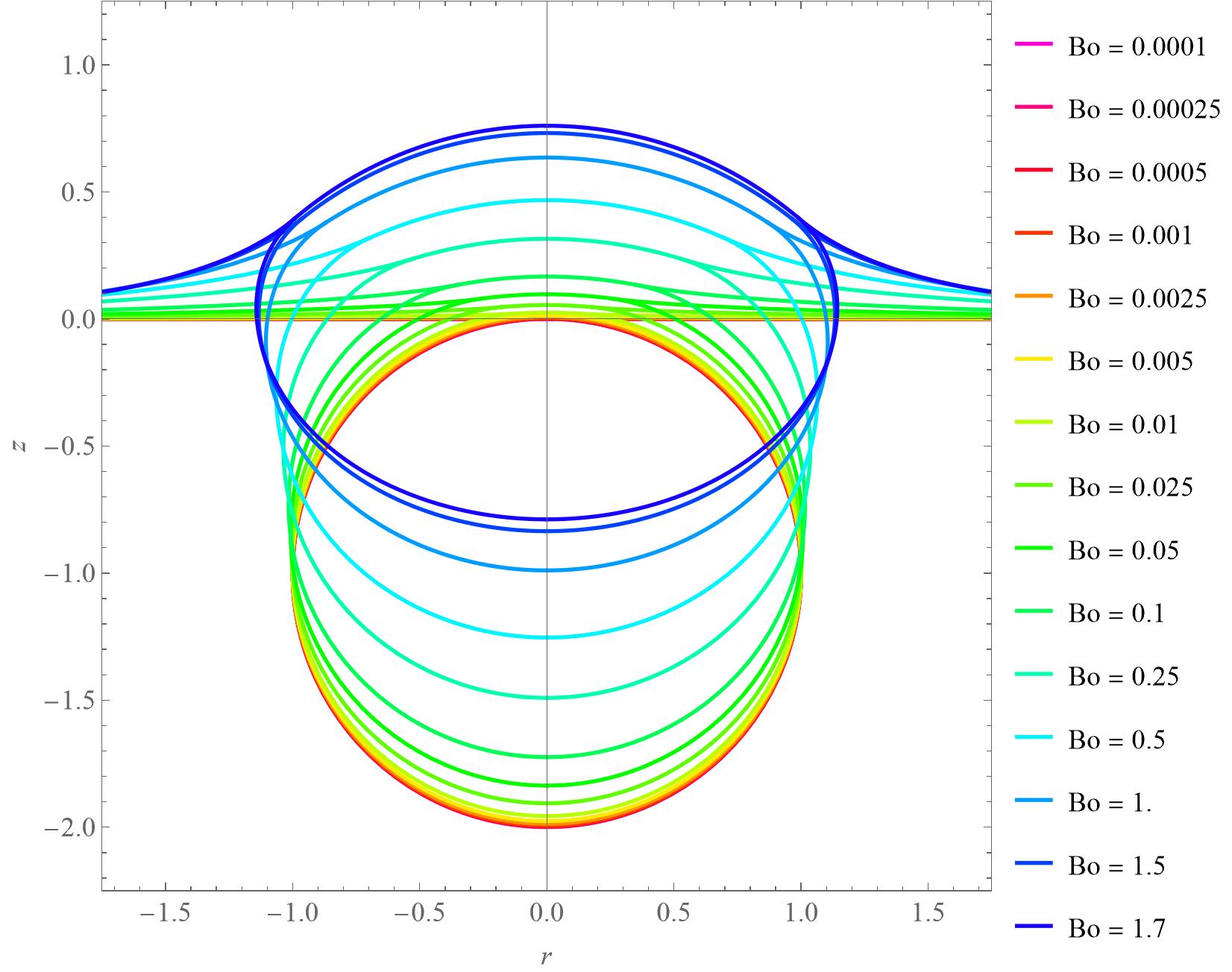}}
  \caption{Equilibrium bubble shapes for given Bond numbers.}
\label{f3}
\end{figure*}

According to (\ref{timecomparison}), it is therefore expected that the bubble will have sufficient time to reach the surface and attain a near-equilibrium shape.
The equilibrium shapes of bubbles at the interface between seawater and air, as a function of the Bond number, constitutes a classical, well-resolvable problem \cite{Toba1959}. Figure \ref{f3} illustrates those shapes for Bo from 10$^{-4}$ to 1.7.

Calculating the equilibrium volume of the film cap that will eventually burst, retract, and yield aerosol droplets in the process entails the calculation of the molecular-scale interactions taking place between the inner and outer faces of the film \cite{Pugh1996}. Given the electrolytic character of clean seawater and assuming that of other effects associated to the presence of surfactants or complex molecules leading to nonlinear viscous effects are absemt, the meta-stable film thickness $h$ can be calculated as that corresponding to the state where the disjoining pressure \cite{Pugh1996} as a function of the film thickness becomes equal to the Laplace pressure. The corresponding expression to calculate $h(s)$ as a function of the surface coordinate $s$ along the cap profile from the top is given by:
\begin{eqnarray}
\left(\frac{A}{6 \pi h^3}-B_{\rm EDL} e^{-h/\lambda_{\rm EDL}}-B_{\rm H} {\rm e}^{-h/\lambda_{\rm H}}\right)&+&\nonumber \\
\rho g R \left(\frac{4}{P_0} \left(1-\cos \left(\frac{P_0 s}{4}\right)\right)-z_0\right)
&+&\frac{P_0 \sigma}{2 R_o}=0
\label{DP}
\end{eqnarray}
where the left term is the disjoining pressure and the right is the gravity plus the Laplace pressures. Here, the three terms of the disjoining pressure are, from left to right:

1- The contribution of van der Waals forces, where $A\simeq 3.5\times  10^{-20}$J is the Hamaker's constant of seawater,

2- The electrostatic double-layer forces, where
\begin{equation}
B_{\rm EDL}= 64 n_0 k_B T \tanh^2\left( \frac{e \psi_0}{4 k_B T} \right).
\end{equation}
Here, $n_0,T,k_B,e$ and $\psi_0$ are the number concentration of ions, temperature, Boltzmann constant, and the surface potential (in absolute value). Again, for seawater at $T=15.5^o$C, $B_{EDL}\simeq 1.146\times 10^{7}$ Pasc, and the Debye layer thickness $\lambda_{EDL}\simeq 0.86$ nm.

3- The structural hydration forces, where $B_{\rm H}\simeq 5 \times 10^8$ Pasc, and $\lambda_{\rm H}=0.25$ nm.

In (\ref{DP}), $z_0$ is the vertical coordinate of the cap origin with respect to the free surface. For small $\overline{\rm Bo}$ numbers, $z_0\ll R_o$, $P_0\simeq 2$ (nearly spherical bubbles) and the cap radius is a small fraction of $R_o$ that, in first approximation, is given by \cite{Howell1999}:
\begin{equation}
R_c = \left(\frac{\overline{\rm Bo}}{3}\right)^{1/2} R_o,
\label{Rc}
\end{equation}

Equation (\ref{DP}) has two metastable solutions when electrostatic double layer (EDL) or hydration forces dominate. This results in extremely thin black film states that do not reflect light. However, once draining is sufficiently developed and $h$ becomes thin enough, van der Waals forces first dominate, producing a rapid collapse that normally precludes formation of the much thinner metastable film solutions. Here, we use the most conservative calculation, i.e., a {\it thicker} film, corresponding to a constant film thickness $h$ given by the value at which a rapid collapse would start.
\begin{equation}
h\simeq \left(\frac{A R_o}{6\pi\sigma}\right)^{1/3}.
\label{hsim}
\end{equation}
resulting approximately $h\simeq \left(\left(\frac{d_o}{2}\right)^2 R_o\right)^{1/3}$, where $d_o\simeq \left(2 A/(3\pi \sigma)\right)^{1/2}$ is a characteristic distance that, for seawater, happens to be close to the molecular size of $0.3$ nm. Equations (\ref{Rc}) and (\ref{hsim}) finally yield a film volume
\begin{equation}
V_{\rm film}\simeq \frac{\pi}{3\times 2^{2/3}} \left(\frac{d_o}{R_o}\right)^{2/3} R_o^3 \, \overline{Bo}.
\label{Vfilm}
\end{equation}

\subsubsection{Total jet volume}

\begin{figure*}
\centerline{\includegraphics[width=0.80\textwidth]{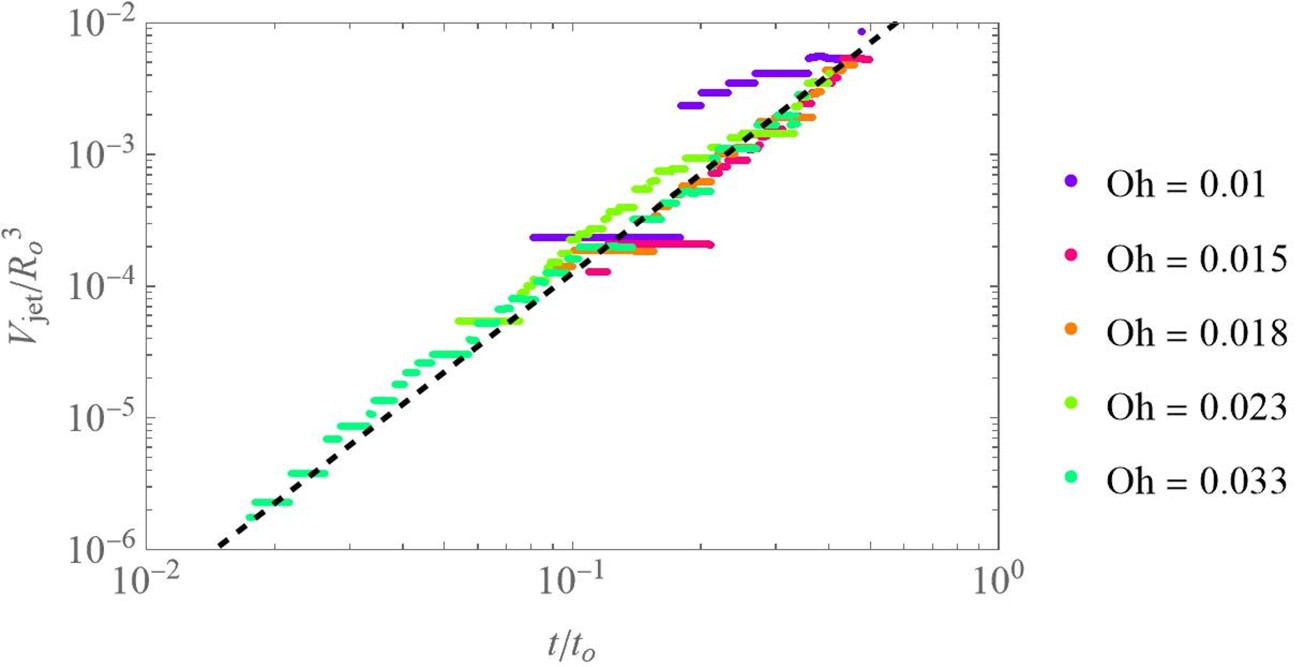}}
  \caption{Ejected volume as a function of time, for various Oh = ${\rm La}^{-1/2}$ values. The total ejected volume results nearly independent of the Laplace number. The slope of the dsahed line is 8/3.}
\label{f4}
\end{figure*}

For sub-millimeter bubble sizes, the range of Laplace numbers that yields emission is from about ${\rm La}\simeq 5\times 10^2$ to about $5\times 10^4$. These values suggest that viscous dissipation would be restricted to regions near the water-air interface, and that there would be a self-similar geometry of the ejection along time, whose scaling was described in \cite{ZKFL00,GL21}. In this case, the scaling was $r \sim t^{2/3}$, which suggests that the instantaneous ejected volume $V_j$ scales as:
\begin{equation}
V_{j}/R_o^3 \sim \int_0^\chi \chi'^3 {\rm d}\chi' \sim \int_0^\chi \chi'^3 \frac{{\rm d}\chi}{{\rm d}\tau} {\rm d}\tau \sim \int_0^\tau \tau'^{5/3} {\rm d} \tau' \sim \tau^{8/3},
\label{Vjet}
\end{equation}
where $\chi=r/R_o$, $\tau=t/t_o$, $t_o=\left(\frac{\rho R_o^3}{\sigma}\right)^{1/2}$ and $r$ is the instantaneous characteristic length scale of the jet being emitted. This scaling is demonstrated using the data from \cite{LG24} in Figure \ref{f4}: The plot shows the total volume ejected in the form of droplets as a function of time.

Figure \ref{f4} demonstrates that the end of ejection occurs systematically around $\tau_{end} \simeq 0.5$, with a value
\begin{equation}
V_j(\tau_{end})=V_{\rm jet} \simeq 6 \times 10^{-3} R_o^3
\end{equation}
for the La numbers of interest here.


\subsubsection{Film versus jet volume: The contribution of film droplets is negligible versus jet droplets in the sub-micron range}

The ratio $V_{\rm film}/V_{\rm jet}$, given by (\ref{Vfilm}) and (\ref{Vjet}), reads
\begin{equation}
V_{\rm film}/V_{\rm jet}\simeq 1.1\times 10^2 \left(\frac{d_o}{R_o}\right)^{2/3}\overline{\rm Bo}.
\label{Vratio}
\end{equation}
To have an idea of how small this ratio is, even for large local accelerations of the bubble, consider a relatively large $\overline{\rm Bo}\sim 0.1$ with the smallest bubble size of interest (${\rm La}\simeq {\rm La}_{cr}=10^3$) i.e. $R_o\simeq 20 \, \mu$m. Since $d_o\simeq 0.3$ nm for water molecules, this yields $V_{\rm film}/V_{\rm jet}\simeq 6.7 \times 10^{-3}$.

\begin{figure*}
\centerline{\includegraphics[width=0.8\textwidth]{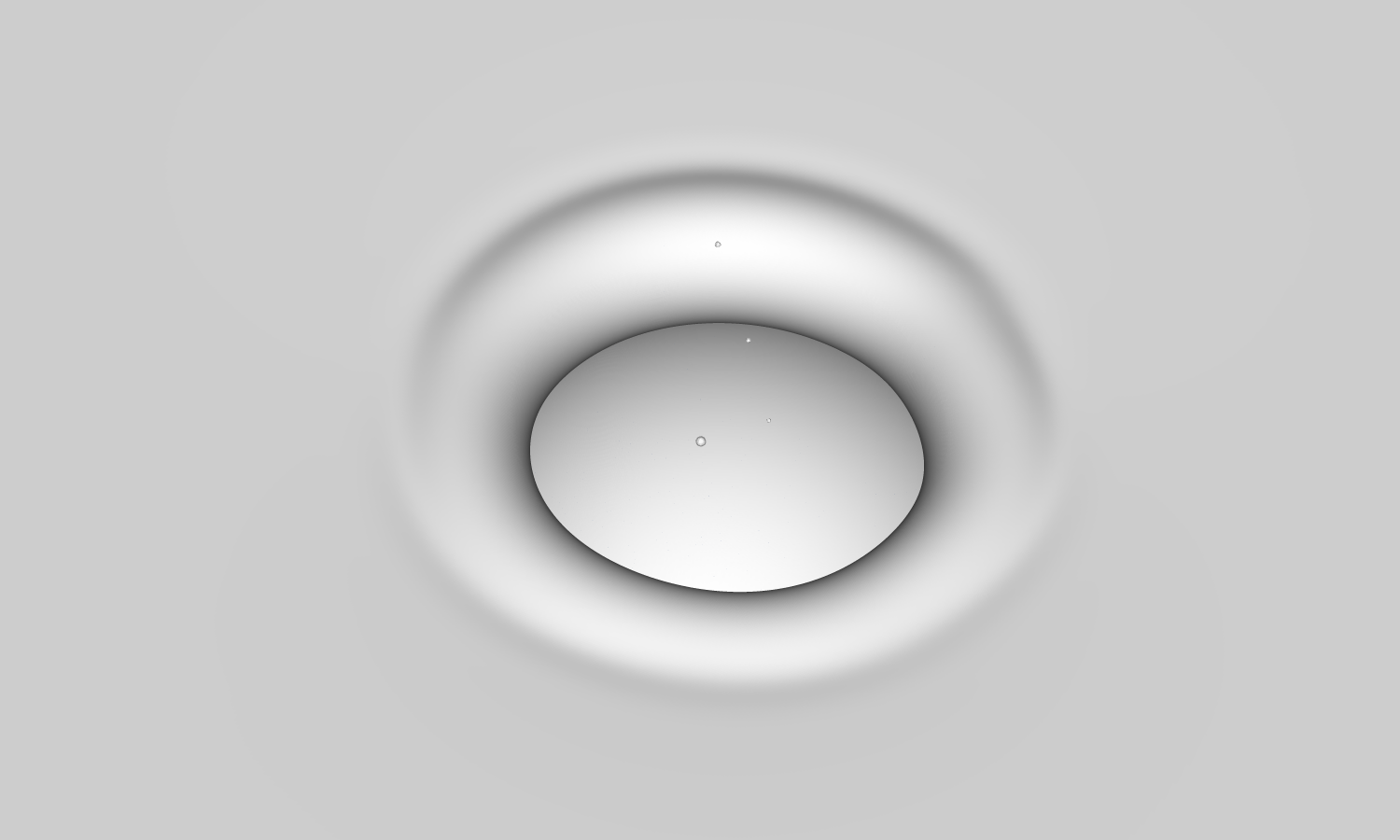}}
\centerline{\includegraphics[width=0.8\textwidth]{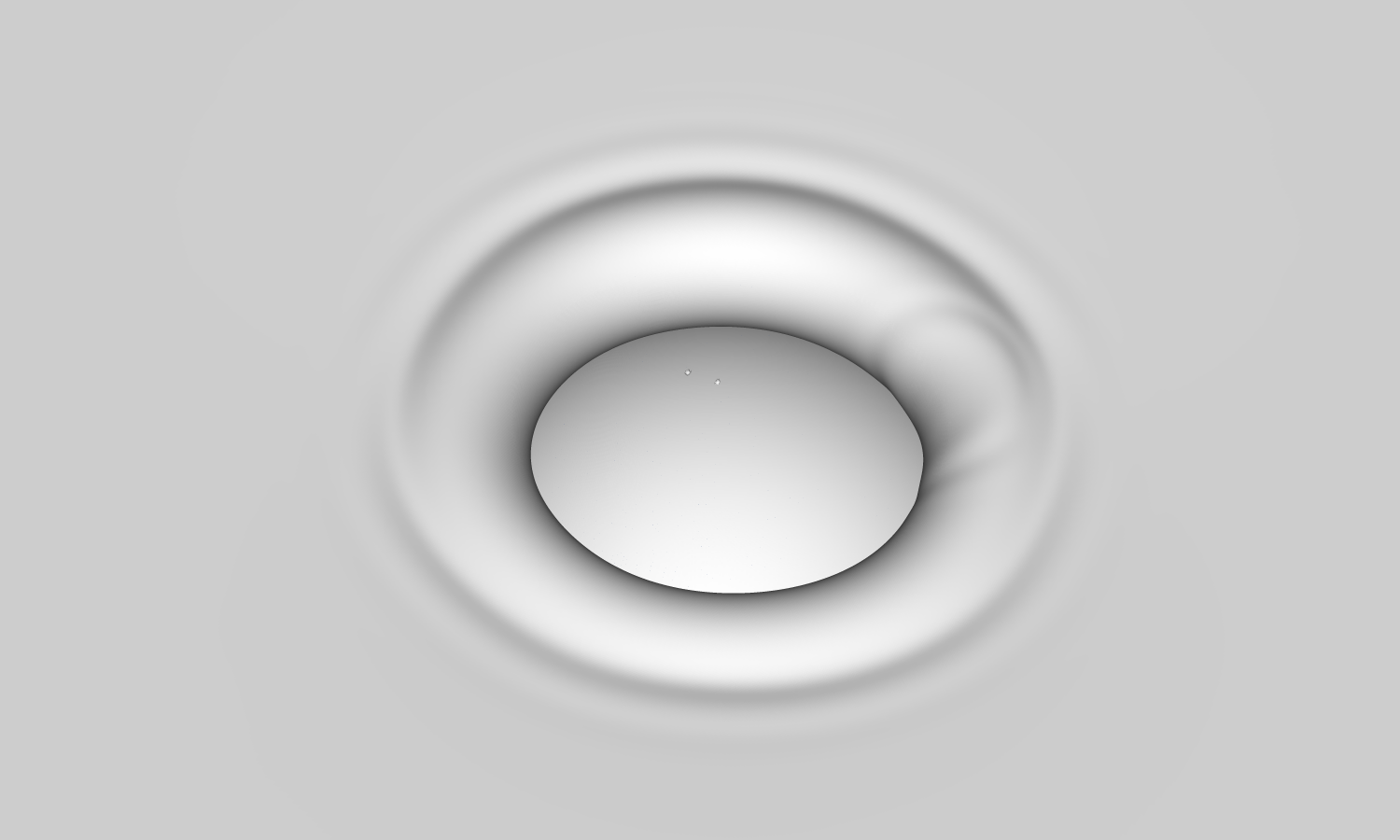}}
\centerline{\includegraphics[width=0.8\textwidth]{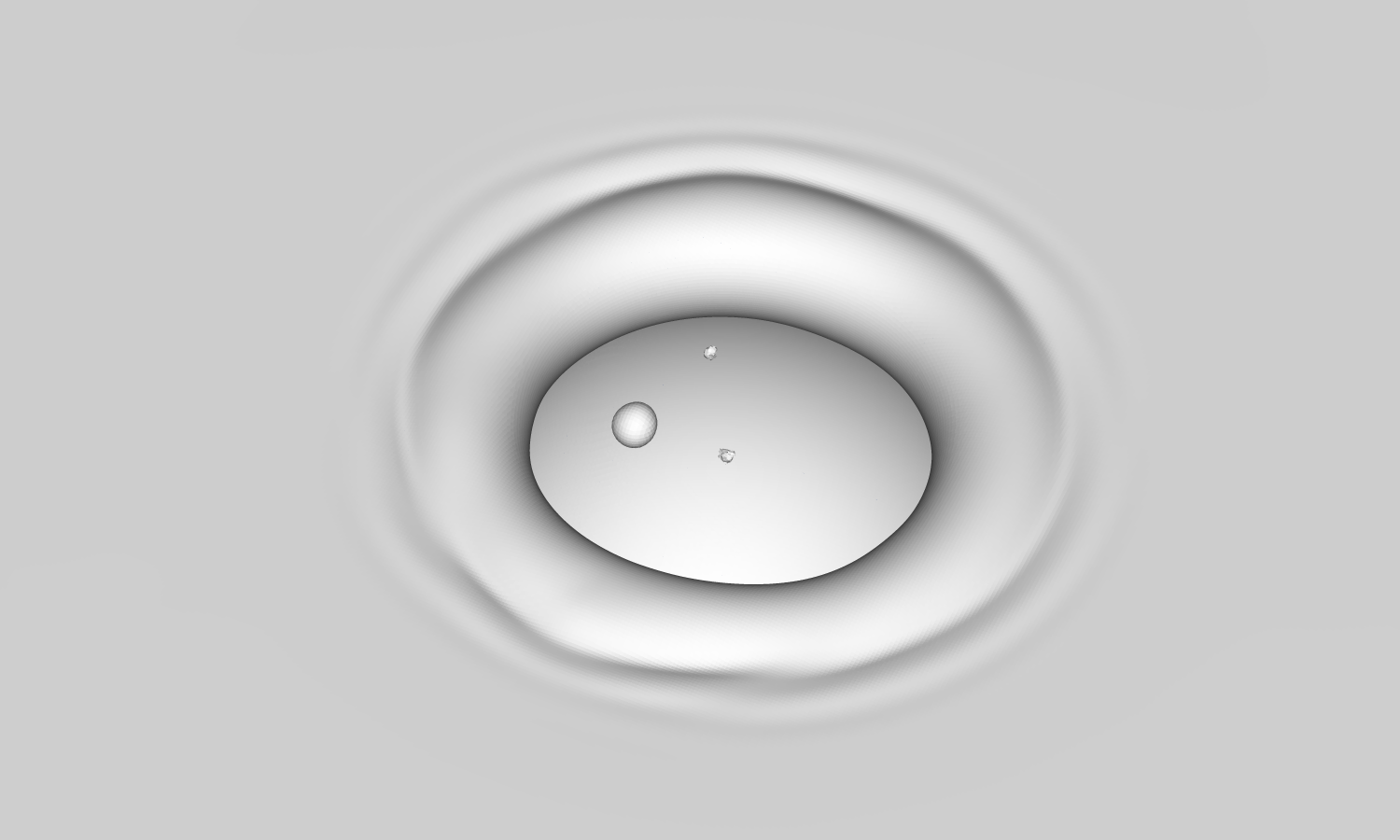}}
  \caption{Instantaneous snapshots of the bubble rim and tiny droplets formed at $t=0.1\, t_\sigma$ after the cap puncture. From top to bottom: bubbles of radii $R_o=20$, 100 and 500 $\mu$m; seawater at 15.5$^o$C. Basilisk \cite{Basilisk} 3D simulations with a level L11 octree grid refinement.}
\label{f41}
\end{figure*}

The equation (\ref{Vratio}) is a fundamental result. It would be unreasonable to think that film breakup could shatter into droplets comparable in size to either $l_\mu$ or the film thickness $h$. Neither experiments nor numerical simulations (see Figure \ref{f41}) in the range of interest support this idea. Even in numerical simulations that produce large local perturbations due to numerical inaccuracies in the bubble's ultimate approach to the surface -- which could be assimilated to the presence of pollutants in the surface microlayer or the discrete impact of air molecules -- the number of droplets produced by the disrupted film breakup is limited. (see Figure \ref{f41}). In most cases, these droplets retract and collapse on the retracting rim. In fact, within the La range of interest, the rim retracts nearly symmetrically because there is no time for azimuthal instabilities to develop and yield fingers, let alone to allow film ``flapping'' due to gas-liquid interaction as proposed in \cite{JRVW22,VWD22}: If any scales were to develop, they would be comparable in size to $l_\mu$. Yet, these scales would be much smaller than the molecular mean free path of air, preventing the finer-grain air-liquid interactions necessary to the development of this mechanism. Thus, only a few nanodroplets could be observed immediately after film puncture (see Figure \ref{f41}).

At most, one could argue that the film and jet breakup mechanisms yield similar probability density functions (PDFs), which irrefutably leads to the conclusion that film droplets contribute negligibly to the sub-micron spray size.

From now on, this study will focus on jet droplets alone.

\subsection{Bubble bursting droplets statistics}

\subsubsection{The number of ejected droplets}

\begin{figure*}
\centerline{\includegraphics[width=0.85\textwidth]{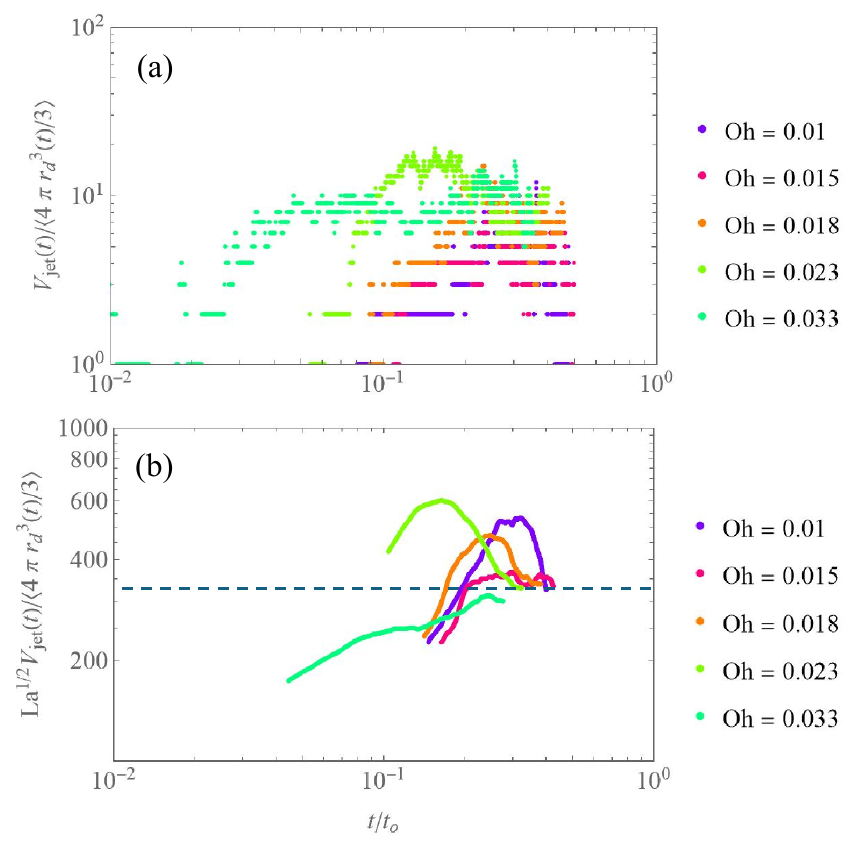}}
  \caption{Instantaneous number of droplets in flight, as a function of time, for various Oh numbers. (a) Instantaneous values. (b) Moving time-averaged values with 100 consecutive points.}
\label{f5}
\end{figure*}

\begin{figure*}
\centerline{\includegraphics[width=0.85\textwidth]{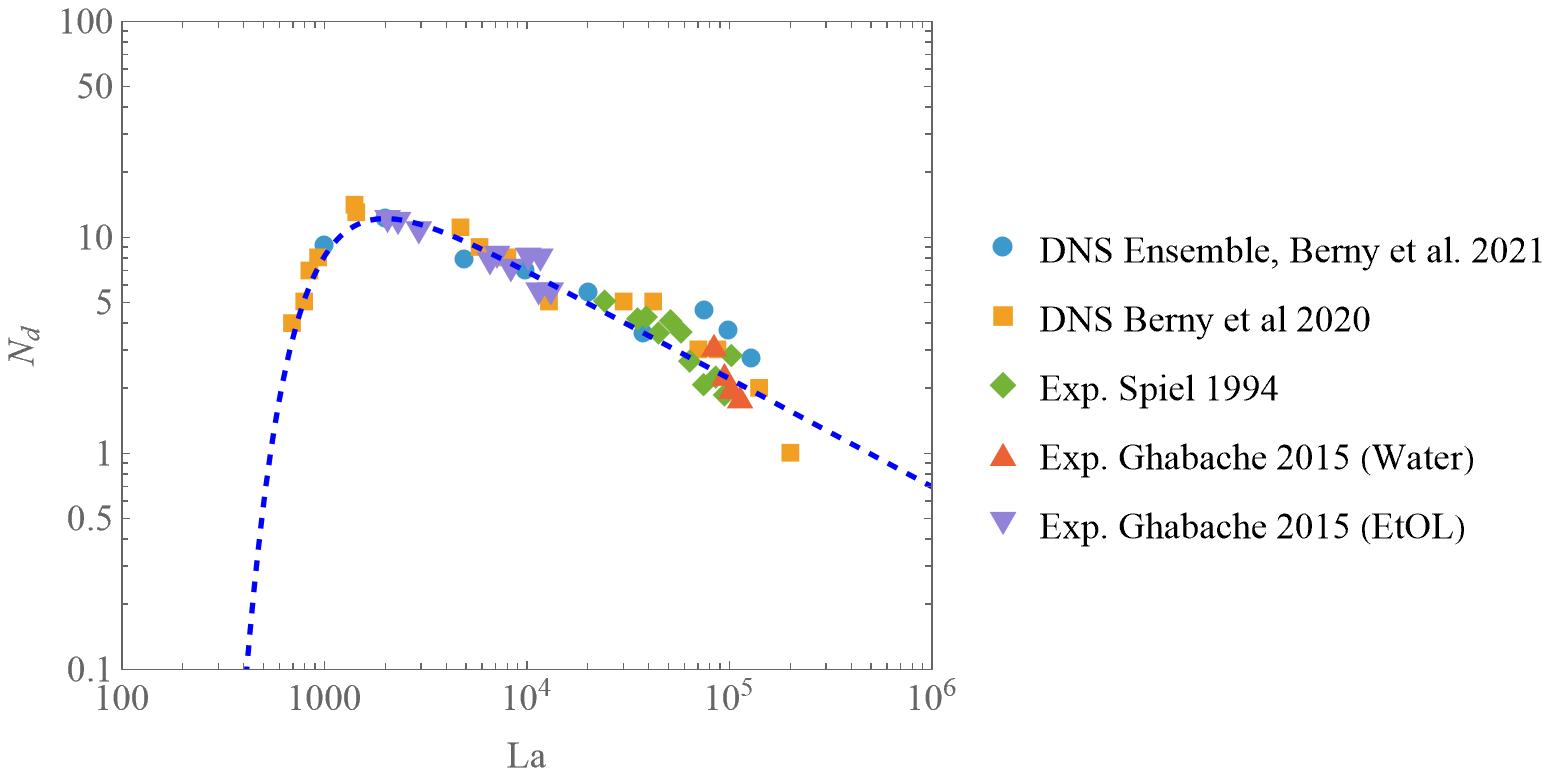}}
  \caption{Number of droplets ejected as a function of La.}
\label{f6}
\end{figure*}

The volume $V_{\rm jet}$ can be expressed in terms of the average number of ejected droplets $N_d$ and the volume mean radius $\langle r_d \rangle_{vol}$ as:
\begin{equation}
V_{\rm jet}=N_d \frac{4}{3}\pi \langle r_d \rangle_{vol}^3
\end{equation}
Figure \ref{f5} gives the instantaneous and the time-averaged number of droplets in flight (moving average with consecutive points 100) during ejection. Figure \ref{f5} suggests that $N_d$ scales as ${\rm La}^{-1/2}$, except for ${\rm Oh} = {\rm La}^{1/2}$ numbers except the one close to the critical value.
This result agrees remarkably well with the published experimental evidences, as shown in Figure \ref{f6}. The dashed black line corresponds to the fitting:
\begin{equation}
N_d(La)\simeq 700\, {\rm La}^{-1/2} \exp\left(-{\rm La}_{\rm cr}/{\rm La}\right).
\label{Nd}
\end{equation}
The resulting volume mean radius of the ejected droplets is:
\begin{equation}
\langle r_d \rangle_{vol} \simeq 1.27\times 10^{-2} {\rm La}^{1/6} {\rm e}^{{\rm La}_{\rm cr}/(3 {\rm La})}
\label{VMR}
\end{equation}
For example, for ${\rm La}=10^4,\, 10^5$ and $10^6$, one has $\langle r_d \rangle_{vol}/R_o \simeq 0.061,\, 0.087$ and 0.127, in agreement with the literature \cite{Tedesco1979,Blanchard1989,Ghabache2016,DGLDZPS18}.

\subsubsection{PDF of jet droplets}

\begin{figure*}
\centerline{\includegraphics[width=1.0\textwidth]{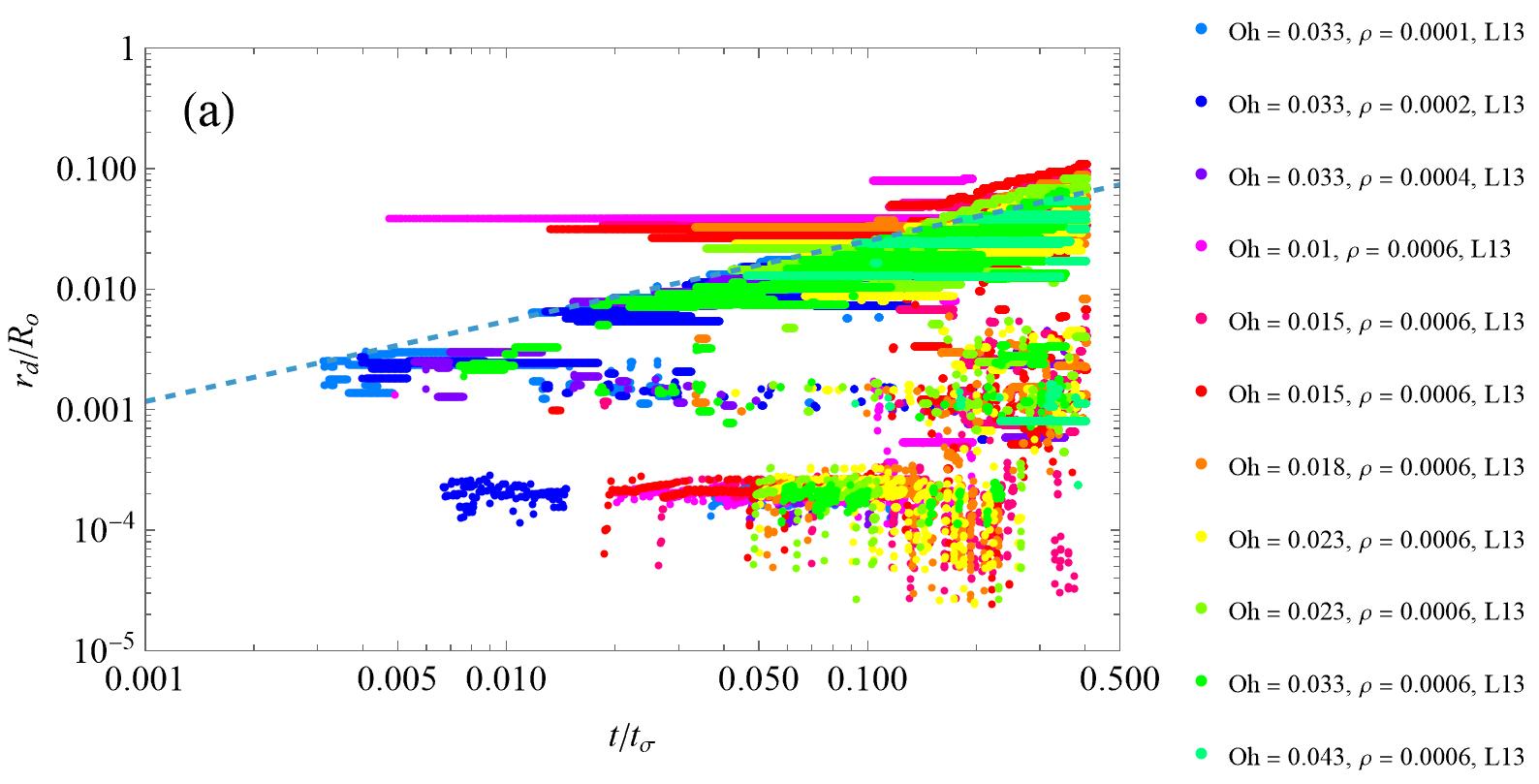}}
\centerline{\includegraphics[width=0.733\textwidth]{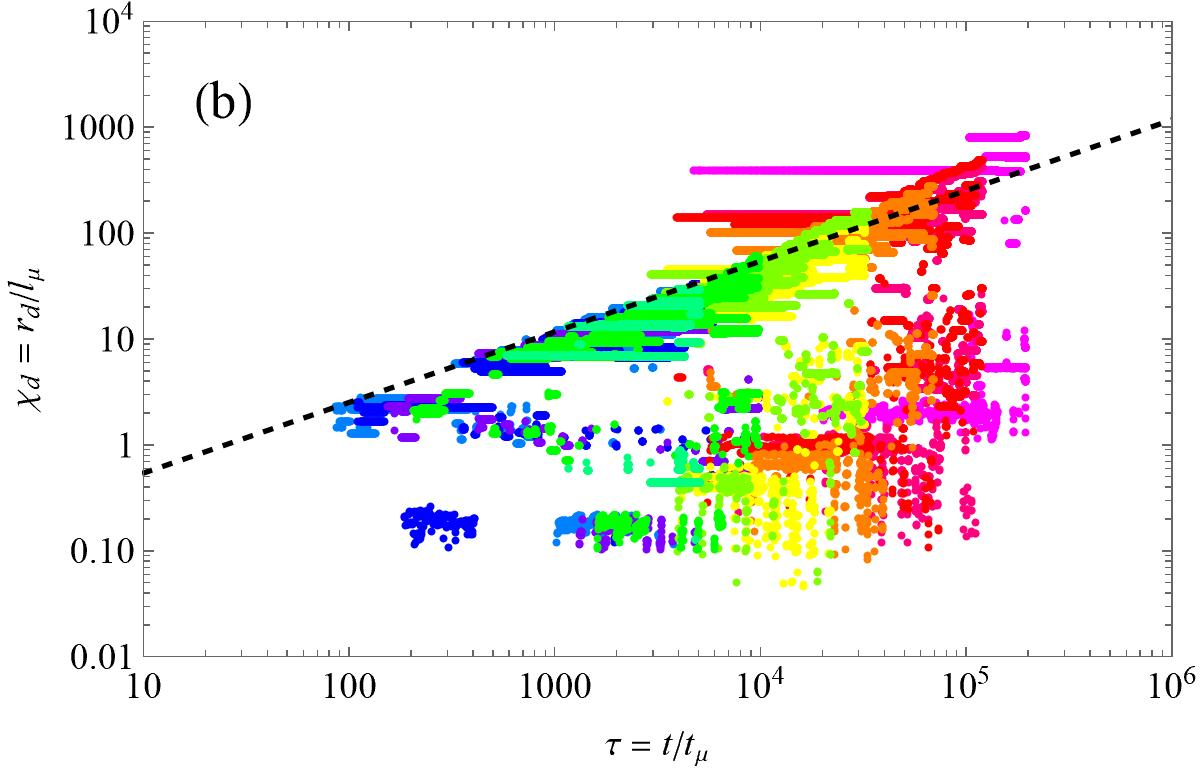}}
  \caption{Instantaneous normalized radii of the droplets in flight during ejection. Radii and times normalized with (a) $R_o$ and $t_\sigma$, and (b) $l_\mu$ and $t_\mu$. Dashed lines indicate the scaling law $\chi_d \sim \tau^{2/3}$ \cite{ZKFL00,GL21}.}
\label{f7}
\end{figure*}

Numerical simulations using Basilisk \cite{Basilisk} in the axisymmetric configuration (Level 13) used in \cite{GL21,LG24} have been performed for Oh numbers from 0.01 to 0.043 (La from 540 to 10$^4$). Figure \ref{f7} shows the instantaneous values of the radii in flight at each instant during ejection, normalized with both $R_o$ and $l_\mu$, as a function of time, normalized with both $t_\mu=\mu^3/(\rho \sigma^2)$ and $t_\sigma=(\rho R_o^3/\sigma)^{1/2}$ (Figures \ref{f7}(a) and (b), respectively).

These figures exhibit key features:

1- A wide range of droplet radii are instantaneously present along the ejection, with values about $0.1 l\mu$ to about $0.1 R_o$.

2- All cases appear to consistently produce droplets in the range around $l_\mu$.

3- The scale of the most frequent droplets ejected at every instant scale as $\chi_d \sim \tau^{2/3}$, consistently with the scaling law already shown in the literature \cite{ZKFL00,GL21}.

\begin{figure*}
\centerline{\includegraphics[width=0.8\textwidth]{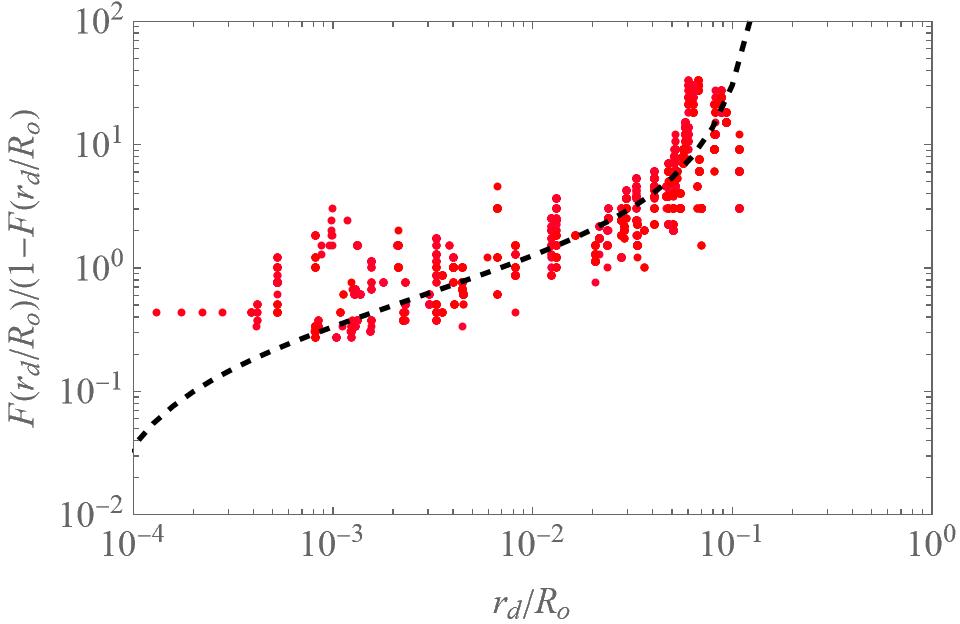}}
  \caption{Discrete form of the function $F/(1-F)\simeq i/(n(t)+1-i)$ obtained from data (La $=1500$; last 0.005$t_\sigma$ time interval of the ejection), and the best fitting (dashed line) using a Generalized Inverse Gaussian model.}
\label{f8}
\end{figure*}

A PDF can be hardly evaluated from the collection of scattered data due to the scarcity of droplets present every instant; however, it is much easier to fit a mathematical form of the cumulative PDF $F=\int_0^{x} f(x'){\rm d} x'$, expressed as $F/(1-F)$, where $F(x_i,t)\simeq i/(n(t)+1)$, and $i$ and $n(t)$ are the index of the $i-$th droplet sorted by size and the number of droplets present in the instant $t$, respectively. This way of finding PDFs was already used in \cite{G23}, since it exhibits the scaling behavior of the two extremes of the distribution with a great level of detail. Figure \ref{f8} represents the function $F/(1-F)\simeq i/(n(t)+1-i)$ for an illustrative case ${\rm La}=1500$, using the droplet population in flight during the last 0.005$t_\sigma$ time interval of the ejection.

An exhaustive exploration of PDF models has been carried out. The best option capturing the cumulative behavior of the distribution of droplets exhibited by the simulations at both smaller and larger droplets extremes has been the {\it Generalized Inverse Gaussian} distribution \cite{Jorgensen1982,Provost2010}, here used with the form:
\begin{equation}
f(x)=\frac{\beta \, x^{\alpha-1} e^{-\frac{1}{2} \left(\left(\frac{x_1}{x}\right)^{\beta}+\left(\frac{x}{x­_2}\right)^{\beta}\right)}}{(2 (x_1 x_2)^{\alpha/2} K_{\alpha/\beta}\left(\left(\frac{x_1}{x­_2}\right)^{\beta/2}\right)},
\label{GIG}
\end{equation}
and $x=r_d/R_o$. The fitting parameters have resulted $\alpha=0.15$, $\beta=4$, and $x_1=0.1 {\rm La}^{-1}$. For consistency with the value of the volume mean $\langle r_d \rangle_{vol}$ given in (\ref{VMR}), the value of $x_2$ should be rigorously obtained from the resolution of the equation
\begin{eqnarray}
\langle r_d \rangle_{vol}&=& \left(\int_{0}^{\infty} x'^3 f(x')dx'\right)^{1/3} = \left(\frac{(x_1 x_2)^{3/2} K_{\frac{\alpha+3}{\beta}}\left(\left(\frac{x_1}{x_2}\right)^{\beta/2}\right)}{K_{\frac{\alpha}{\beta}}\left(\left(\frac{x_1}{x_2}\right)^{\beta/2}\right)}\right)^{1/3}=\nonumber\\
&\simeq & 1.27\times 10^{-2} {\rm La}^{1/6} {\rm e}^{{\rm La}_{\rm cr}/(3 {\rm La})}.
\label{x2}
\end{eqnarray}
However, for the purposes of this work (sub-millimeter bubbles, with La values from $10^3$ to less than $5\times 10^4$), an approximate expression such as $x_2\simeq 0.1(1+{\rm La}/10^4)^{1/2}$ for the solution of (\ref{x2}) provides sufficient accuracy.

\section{Result and discussion: the ocean fine spray size distribution. Comparison with experiments}

\begin{figure*}
\centerline{\includegraphics[width=1.0\textwidth]{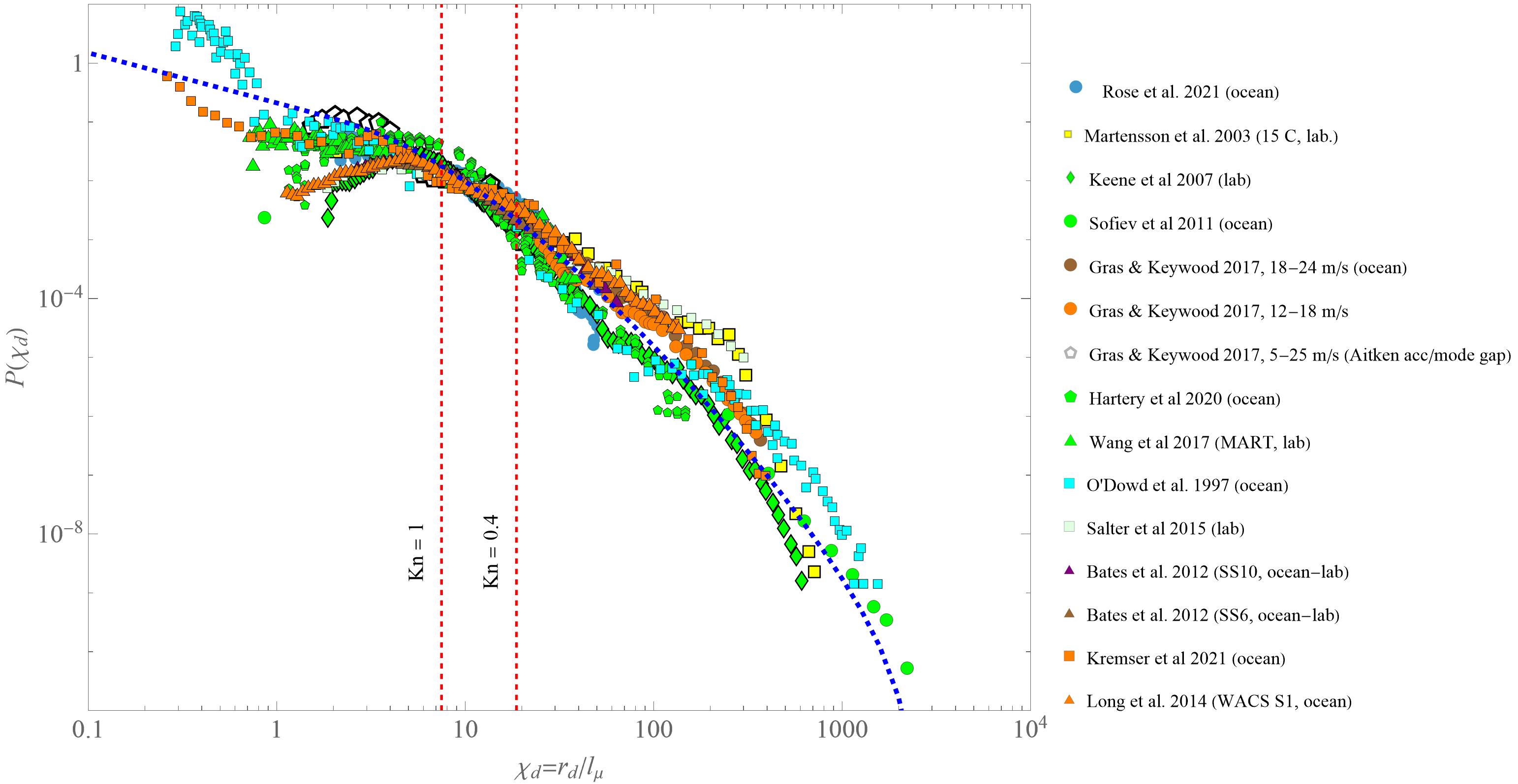}}
  \caption{Data gathered in \cite{G23} from the literature on experimentally measured size distributions of aerosols from the ocean. The dashed blue line is the predicted marginal PDF here obtained, with no fitting parameters except those directly obtained from the numerical simulations, and the isolated information on (i) the bubble size distribution, and (ii) the number of droplets generated by bubble jetting.}
\label{f9}
\end{figure*}

Incorporating the expressions (\ref{exp}), (\ref{Nd}) and (\ref{GIG}) in equation (\ref{marginalPDF}), one obtains the predicted global PDF of droplet sizes produced by the ocean surface in the micron- and sub-micron range. The collection of data gathered in \cite{G23} is used in Figure (\ref{f9}) to show a comparison between the predicted marginal PDF here obtained and the data.
The predicted PDF models nearly perfectly the collected data in the range of interest of SSA for CCN, INP \cite{Mitts2021}, and atmospheric albedo, from about 25nm to 2.5 $\mu$m salt relics sizes ($r_d/l_\mu=2$ to 200). In this range, the CCN and INP from the salt SSA are expected to keep at least a dominant salt composition, with other component in lesser amount depending on many parameters. Smaller sizes are likely to come from a mix of SMA (probably the dominant population in this range) and aged SSA, and consequently exhibit a large dispersion due to the geographical and seasonal variability of collected data origins \cite{Mayer2020}. In fact, the variability of physicochemical and biochemical processes undergone by these ultrafine particles and molecular clusters in flight is extreme due to their small Knudsen numbers and their extremely large residence times in the turbulent atmosphere.

In the other extreme, the film droplets begin to play a role above 1 $\mu$m. Due to the potentially different composition that these droplets may have due to their different formation mechanisms and differential exposition to the marine surface microlayer \cite{Wang2017,GL21}, their composition is expected to show increased variability.

Moreover, while this work does not entirely exclude the possibility that the film cap could generate a large number of nanodroplets -- particularly via the {\it hen-and-chick} mechanism here described at sufficiently  effective Bond numbers --, solid arguments are presented that point to jetting as the primary source. It is crucial to highlight two points, even though the final comparison with experimental data does not distinguish the origin of the particles:

1- The jetting model, based on numerical simulations, generates a marginal PDF that is quantitatively consistent with experimental measurements without requiring additional assumptions.

2- Dynamically, nanodroplets from jetting are ejected with velocities orders of magnitude higher than those from the film. As observed in simulations, the latter possess such a low residual velocity that they tend to remain near the surface and collapse back onto it.

Therefore, the ability of jet droplets to escape the surface boundary layer and effectively contribute to the atmospheric aerosol population unequivocally establishes this mechanism as the dominant contributor.

Furthermore, the results obtained provide an overall quantification, or global PDF model, and do not differentiate among the numerous parameters affecting the aerosol generation from the ocean surface, such as wind speed and water temperature. However, we hypothesize that these parameters will exert a sub-dominant influence on a normalized PDF, such as the one obtained in this study, and the subsequent data collection appears to support this hypothesis. The rate of production or aerosol flux from the ocean \cite{Veron2015} is a critical variable in this analysis. Once the underlying probability distribution function (PDF) associated with common physical mechanisms (independent of the intensity of winds, in first approximation) is identified, the aerosol flux from the ocean can be more effectively isolated. This enables a more nuanced and accurate characterization of its dependence on parameters such as wind speed, which influences flux through enhanced turbulence and air-water mixing, and temperature, which affects surface tension, viscosity, and the concentration of biological compounds within the marine surface microlayer.

Finally, by first characterizing and properly modelling the generic PDF of SSA generated by well-established mechanisms, the marginal PDFs of re-suspended microorganisms and micro-/nanoplastics \cite{Metal21} can be more readily derived, establishing a quantitative framework that enhances understanding and prediction of their environmental impact and atmospheric dispersion pathways.

\section{Declarations}

\begin{itemize}
\item Funding: Partial funding from the the Spanish Ministry of Science and Innovation, grant no. PID2022-14095OB-C21.
\item Conflict of interest/Competing interests (check journal-specific guidelines for which heading to use): The author declares no conflicts of interest.
\item Ethics approval: Not applicable
\item Consent to participate: Not applicable
\item Consent for publication: Not applicable
\item Availability of data and materials: Data and materials can be made available upon reasonable request.
\item Code availability: Not applicable
\item Authors' contributions: JMLH: Basilisk simulations, data collection and curation. MAH: Numerical model: calculations, and funding. AMGC: paper conception; analytic model: conception, calculations; data analysis, data collection; writing and funding.
\end{itemize}


\end{document}